\documentclass[12pt]{article}

\usepackage{amsmath}
\begin{document}
\vspace*{-.6in}
\thispagestyle{empty}
\begin{flushright}
CALT-68-2330\\
hep-th/0105253
\end{flushright}

\vspace{.5in}

{\Large
\begin{center}

On the Gauge Invariance of the Chern-Simons \\
 Action for $N$ D-Branes
\end{center}}
\begin{center}
Calin Ciocarlie\footnote{email: calin@theory.caltech.edu} \\
\emph{ California Institute of Technology, Pasadena, CA  91125 USA}
\end{center}
\vspace*{1in}

\begin{center}
\textbf{Abstract}
\end{center}
\begin{quotation}
\noindent In this short note we provide a proof that the Chern-Simons part of the action for $N$ D-branes is invariant under gauge transformations of the RR fields of the type $C_p \rightarrow C_p+d\Lambda_{p-1}$, and rewrite the action in a form that makes this symmetry manifest.
\end{quotation}
\vfil
\newpage

\section{Introduction}

   Using the principle of consistency under T-duality transformation, the authors of ~\cite{myers, taylor1} extended the Chern-Simons part of the action for a $D_{p-1}$ brane to the case of $N$ coincident $D_{p-1}$ branes.   The extended action contains extra terms that, in general, give a non-trivial coupling between the $N$ $D_{p-1}$ branes and a higher rank RR form, $C_{p+2k}$. As mentioned in ~\cite{schwarz}, it is not obvious whether the extended action is still invariant under gauge transformations of the type,  $ C_p \rightarrow C_p+d\Lambda_{p-1}$. It is the purpose of this note to investigate this question. It was not clear, a priori, whether to expect this to work. The fact that it does seems quite remarkable.

   The world-volume action for the $D_{p-1}$ branes will be written in the static gauge: one can use spacetime diffeomorphisms to define the fiducial world-volume as $x^i=0$, $i=p,\ldots,9$, and world-volume diffeomorphisms to match the coordinates of the branes with the remaining space-time coordinates, i.e. $\sigma^a=x^a$, $a=0,\ldots,p-1$. The transverse displacements of the branes are  $\Delta x^i=(2 \pi \alpha')\phi^i \equiv \lambda \phi^i$, where $\phi^i$ is an $N \times N$ matrix.

   The Chern-Simons term for $N$ coincident $D_{p-1}$ branes is given by ~\cite{myers},
\begin{equation}
S_{CS}=\mu_{p-1}\int Str\left(P \left[ e^{i\lambda i_\phi i_\phi} \left(\Sigma C^{\left(n\right)}e^B\right)\right]e^{\lambda F}\right),
\end{equation}
   $P(\ldots )$ represents the pullback, 
   $i_\phi i_\phi$ defines an inner product, e.g. $i_\phi i_\phi C^{(2)} = \frac{1}{2}\left[\phi^j,\phi^i\right]C^{(2)}_{ij},$
   $F_{ab}$ is the gauge field strength living on the D-brane, and
   $\sigma$'s are the coordinates parallel to the directions of the branes.

    In this action, the background fields are considered to be functionals of the non-abelian scalars $\phi$'s, as suggested in ~\cite{douglas}, while the pull-backs are defined in terms of covariant derivatives, $D_a\phi^i$, as in ~\cite{hull}. Furthermore the action includes a symmetrized trace prescription: we have to take  a symmetrized average over all orderings of $\phi^i$, $D_a\phi^i$, $F_{ab}$, and pairs of $\underline{\phi^{2k'}\phi^{2k'-1}}$ from the inner product. This prescription is in agreement with results obtained in ~\cite{taylor2} from matrix theory considerations.  
    For simplicity, the gauge field living on the brane ($F_{ab}$) and the background $NS-NS$ field $B$, are initially taken to be 0. Even for this simplified case, the demonstration of gauge invariance is rather long and subtle. We have tried to make it as clear and simple as possible. Later on we will generalize the proof to non-zero $F$. Gauge invariant expressions will be given in equation (33), for $F=0$, and in (39) for $F\neq0$. At the end, we will also explain how to include a $B$ field.

\section{ Gauge invariance,  $F=0$ case}
     In the following, we will show that the coupling between $N$ $D_{p-1}$ branes and a $C_{p+2k}$ RR form is invariant under the transformation $C\rightarrow C+d\Lambda$. A particular case of this problem was proved in [6], in a matrix
 theory context, working in the momentum basis. Here, we generalize, considering $D_{p-1}$ branes instead of $D_0$ branes, with non-trivial pull-back and $F_{ab}$  terms. (Nonzero $F$ will be considered in the next section.)

Specializing equation (1) to the case $F=B=0$, the coupling between a $C_{p+2k}$ RR form and $N$ $D_{p-1}$ branes is given by:
\begin{equation}
 \mu_{p-1}\int Str\left(P\left[\frac{\left(i\lambda i_\phi i_\phi \right)^k}{k!} C_{p+2k} \right]\right)
\end{equation}

Each of the  RR fields $C_{p+2k}$ are functionals of the transverse coordinates $\phi$ :
\begin{equation}
C(\sigma,\phi) = e^{\lambda\phi^i\partial_{x^i}} C^0(\sigma,x^i)|_{x^i=0}=\sum_{n,{i_n}} \frac {\lambda^n}{n!} \phi^{i_1}\ldots \phi^{i_n} \partial_{x^{i_1}}\ldots \partial_{x^{i_n}} C^0(\sigma,x^i)|_{x^i=0}
\end{equation}
where $C^0(\sigma,x^i)$ is the background RR field.
If $\lambda \phi^i$ are the transverse displacements of the branes, the pullback of a $p$  form, $\Omega_p$, in the static gauge is:
\begin{equation}
[P(\Omega_p)]_{a_1\ldots a_p}= \Omega_{\mu_1\ldots \mu_p}\left(\delta_{a_1}^{\mu_1}I_N+\lambda\frac{\partial\phi^{\mu_1}}{\partial\sigma^{a_1}}\right)\ldots \left(\delta_{a_p}^{\mu_p}I_N+\lambda\frac{\partial\phi^{\mu_p}}{\partial\sigma^{a_p}}\right)
\end{equation}
where $I_N$  is an $N \times N$ unit matrix, and $\Omega_p$ should be considered a functional of the $\phi$'s.
The indices  $\mu$'s run over all coordinates, so we will take $\phi^\mu=0$ for the $\mu$'s parallel to the direction of the branes. As defined in the previous equation the pullback of an antisymmetric form is not necessarily an antisymmetric form since, as $N \times N$ matrices, $\partial_{a_i}\phi^i$ do not commute in general. However, as part of the symmetrized trace prescription we should take a symmetrized average over all orderings of $\partial_{a_i}\phi^i$, thus enforcing antisymmetry on  the $a$'s. \newline
 With antisymmetry enforced  on the $a$'s, (4) becomes: 

$[P(\Omega_p)]_{[a_1\ldots a_p]} = \Omega_{a_1\ldots a_p}+\lambda p\Omega_{i_1[a_2\ldots a_p}\partial_{a_1]}\phi^{i_1}+\ldots +\lambda^l\frac{p!}{l!(p-l)!}\Omega_{i_1\ldots i_l[a_{l+1}\ldots a_p}\partial_{a_1}\phi^{i_1}\ldots \partial_{a_l]}\phi^{i_l}$
\begin{equation}
+\ldots +\lambda^p \Omega_{i_1\ldots i_p}\partial_{[a_1}\phi^{i_1}\ldots \partial_{a_p]}\phi^{i_p}.
\end{equation}
We are going to use this equation for  $\Omega_p\equiv\frac{\left(i\lambda i_\phi i_\phi \right)^k}{k!} C_{p+2k}$.
Combining equation (3) and (5), one gets the $C_{p+2k}$ coupling of $N$ $D_{p-1}$ branes (for $F_{ab}=0$)  as :
\begin{equation}
\sum_{l,n} \frac{\mu_{p-1}\lambda^{k+n+l}i^kp!}{k!n!l!(p-l)!}\partial_{x^{i_1}}\ldots \partial_{x^{i_n}} C^0_{i_1'\ldots i_{2k}'j_1\ldots j_l[a_{l+1}\ldots a_p} Str \left(\partial_{a_1}\phi^{j_1}\ldots \partial_{a_l]}\phi^{j_l}\phi^{i_1}\ldots \phi^{i_n}\phi^{i_{2k}'}\phi^{i_{2k-1}'}\ldots \right)
\end{equation}
where   $0\le l\le p.$

Notice that the $Str\left(\ldots \right)$ expression involves symmetrizing over all the $\partial_{a_s}\phi^{j_{s}}$, for $s=1,2,\ldots ,l$,  also over all the $\phi^{i_q}$, for $q=1,2,\ldots n$, and all the pairs $\phi^{i_{2j}'}\phi^{i_{2j-1}'}$, for $j=1,2\ldots k$. 
We can rewrite this term as $\mu_{p-1}\sum_{l,n}\lambda^{k+n+l} b^n_l$, where 
\begin{center}
$b^n_l=\frac{i^k p!}{k!n!l!(p-l)!}\left(\partial_{x^{i_1}}\ldots \partial_{x^{i_n}}\right) C^0_{i_1'\ldots i_{2k}'j_1\ldots j_l[a_{l+1}\ldots a_p} Str \left(\partial_{a_1}\phi^{j_1}\ldots \partial_{a_l]}\phi^{j_l}\phi^{i_1}\ldots \phi^{i_n}\phi^{i_{2k}'}\phi^{i_{2k-1}'}\ldots \right)$ 
\end{center}   
In the previous equation we antisymmetrized over all the $a$'s, and this will be implicit in the rest of this note. 

In order to show that the coupling is invariant, up to a total derivative,  under gauge transformations $C\rightarrow C+d\Lambda$, we will try to write $\sum_{l,n} b^n_l$ as a sum of total derivatives  and gauge invariant  terms that depend on  the field strength of the RR field.  Integrating $b^n_{l>0}$ by parts with respect to $\sigma^{a_l}$, and  dropping the resulting total derivatives and field strength terms, we can express $b^n_{l>0}$ as a sum of two types of terms. (we will keep track of the field strength terms  and will present them  later.)  The first type of term for $b^n_l$ will cancel against the second type of term in the expansion for $b^{n+1}_{l-1}$. In this way all the terms cancel, except  for the first term in $b^n_{l=1}$ and $b^n_{l=0}$. (The second term of $b^n_l$ will turn out to be 0 for $l=l_{max}=p$, or for $n=n_{min}=0$.)  
   
When integrating $b^n_l$ by parts with respect to $\sigma^{a_l}$, we will get terms in which $\partial_{a_l}$ acts either outside the trace on $C^0$, or inside on $\phi$'s. For the part inside the trace, for simplicity of notation, we will only write down the $\phi$ terms that have changed  after integration by parts. 
Note that due to the antisymmetry in the $a$'s, $\partial_{a_l}\partial_{a_s}\phi^j \rightarrow 0 $. 
Let's denote by $U^n_l$ the factor outside the trace, 
\begin{center}
$U^n_l =\frac{i^kp!}{k!n!l!(p-l)!}\left(\partial_{x^{i_1}}\ldots \partial_{x^{i_n}}\right) C^0_{i_1'\ldots i_{2k}'j_1\ldots j_l[a_{l+1}\ldots a_p}.$ 
\end{center}

With these conventions, dropping the total derivative part,
\begin{equation}
\begin{split}
b^n_l
&= (U^n_l) Str\left(\ldots \partial_{a_l}\phi^{j_l}\ldots \right)= (-\partial_{a_l})(U^n_l)Str\left(\ldots \phi^{j_l}\ldots \right)  \\
&- k\left[(U^n_l) Str\left(\ldots \phi^{j_l}\ldots \partial_{a_l}\phi^{i_{2k}'}\phi^{i_{2k-1}'}\ldots \right)+(U^n_l)Str\left(\ldots \phi^{j_l}\ldots \phi^{i_{2k}'}\partial_{a_l}\phi^{i_{2k-1}'}\ldots \right)  \right] \\
& - n (U^n_l)  Str\left(\ldots \phi^{j_l}\ldots \partial_{a_l}\phi^{i_n}\ldots \right).
\end{split}
\end{equation}
The factor of $k$ comes from the $k$ pairs of $\phi^{j'}\phi^{j'-1}$ of the inner product and $n$ from the $n$ $\phi$'s of the Taylor series expansion of the RR form. 
Let:
\begin{center}

$A_1  =  (U^n_l) Str\left(\ldots \phi^{j_l}\ldots \partial_{a_l}\phi^{i_{2k}'}\phi^{i_{2k-1}'}\ldots \right) $ 

$A_2  =  (U^n_l) Str\left(\ldots \phi^{j_l}\ldots \phi^{i_{2k}'}\partial_{a_l}\phi^{i_{2k-1}'}\ldots \right) $ 

$D^n_l     =  (U^n_l) Str\left(\ldots \phi^{j_l}\ldots \partial_{a_l}\phi^{i_n}\ldots \right)$.
\end{center}

\begin{equation}
b^n_l=(-\partial_{a_l})(U^n_l)Str\left(\ldots \phi^{j_l}\ldots \right)-k(A_1+A_2)-nD^n_l.
\end{equation}

Writing,

$\partial_{x^{i_n}}C^0_{i_1'\ldots i_{2k}'j_1\ldots j_la_{l+1}\ldots a_p}=(p+2k+1)\partial_{[x^{i_n}}C^0_{i_1'\ldots i_{2k}'j_1\ldots j_la_{l+1}\ldots a_p]}+\partial_{x^{i_1'}}C^0_{i_ni_2'\ldots a_p}+$ [$(p+2k-1)$ more terms obtained interchanging $i_n$ with all the other indices], \newline
we can rewrite $D^n_l$ as:

$D^n_l$= (gauge invariant term)+
\begin{equation}
    + D^n_l|_{i_n\leftrightarrow i_1'}+\ldots +D^n_l|_{i_n\leftrightarrow i_{2k}'}+D^n_l|_{i_n\leftrightarrow j_1}+\ldots +D^n_l|_{i_n\leftrightarrow j_l}+D^n_l|_{i_n\leftrightarrow a_{l+1}}+\ldots +D^n_l|_{i_n\leftrightarrow a_p}
\end{equation}
(notation: $D^n_l|_{i_n\leftrightarrow i_s}$ means  that in the term  outside the symmetrized trace, i.e. $U^n_l$,  we interchange $i_n$ with $i_{s}$, and if these are dummy indices, this is equivalent to keeping the outside term $U^n_l$ the same while interchanging $i_n$ with $i_s$ inside the trace.) Note that  every time we interchange two indices from the set of indices $i'$ and $j$ inside the trace we get a minus sign since these indices are contracted with indices of the RR form in the term ouside the trace.

Since,\newline
$\hspace*{.5cm}D^n_l=\frac{i^kp!}{k!n!l!(p-l)!}\left(\partial_{x^{i_1}}\ldots \partial_{x^{i_n}}\right) C^0_{i_1'\ldots i_{2k}'j_1\ldots j_l[a_{l+1}\ldots a_p} Str\left(\ldots \phi^{j_l}\ldots \partial_{a_l]}\phi^{i_n}\ldots \right)$ \newline
we have the following relations:

$D^n_l|_{i_n\leftrightarrow j_l} = b^n_l$,\newline
the original term. 

$D^n_l|_{i_n\leftrightarrow j_s}=(U^n_l)Str\left(\ldots \phi^{j_l}\ldots \partial_{a_l}\phi^{j_s}\ldots \partial_{a_s}\phi^{i_n}\ldots \right)$
\begin{center}
$=-(U^n_l)Str\left(\ldots \phi^{j_l}\ldots \partial_{a_s}\phi^{j_s}\ldots \partial_{a_l}\phi^{i_n}\ldots \right)= -D^n_l$,
\end{center}
 for any $s=1,\ldots l-1$, from the antisymmetry in the $a$'s.

$D^n_l|_{i_n\leftrightarrow i_{2k}' }+D^n_l|_{i_n\leftrightarrow i_{2k-1}'}=D^n_l|_{i_n\leftrightarrow i_{2j}'}+D^n_l|_{i_n\leftrightarrow i_{2j-1}'}$

Using the above relations, equation (9) becomes

$D^n_l$= (gauge  invariant term)+
\begin{equation}
+k(D^n_l|_{i_n\leftrightarrow i_{2k}'}+D^n_l|_{i_n\leftrightarrow i_{2k-1}'})-(l-1)D^n_l+b^n_l+D^n_l|_{i_n\leftrightarrow a_{l+1}}+\ldots +D^n_l|_{i_n\leftrightarrow a_p}.
\end{equation}
 Dropping the gauge invariant term, 
\begin{equation}       
  l(D^n_l)=k(B_1+B_2)+b^n_l+ D^n_l|_{i_n\leftrightarrow a_{l+1}}+\ldots +D^n_l|_{i_n\leftrightarrow a_p,}
\end{equation} where we have defined:
\begin{center}
$B_1=D^n_l|_{i_n\leftrightarrow i_{2k}'}=(U^n_l)Str\left(\ldots \phi^{j_l}\ldots \underbrace{\phi^{i_n}\phi^{i_{2k-1}'}}\ldots \partial_{a_l}\phi^{i_{2k}'}\ldots \right)$

$B_2=D^n_l|_{i_n\leftrightarrow i_{2k-1}'}=(U^n_l)Str\left(\ldots \phi^{j_l}\ldots \underbrace{\phi^{i_{2k}'}\phi^{i_n}}\ldots \partial_{a_l}\phi^{i_{2k-1}'}\ldots \right)$
\end{center}
The notation $\underbrace{\phi^i\phi^j}$ means that $\phi^i,\phi^j$ show up together, as one entry, in the symmetrized trace prescription. In this way, the prescription, after interchanging some of the indices inside the trace, is consistent with the initial one.  

Using the last equation to replace the $D^n_l$ term in equation (8), we find:
\begin{equation}
b^n_l=(-\partial_{ a_l})(U^n_l)Str\left(\ldots \phi^{j_l}\ldots \right)-k(A_1+A_2)-\frac{n}{l}[ k(B_1+B_2)+b^n_l+D^n_l|_{i_n\leftrightarrow a_{l+1}}+\ldots +D^n_l|_{i_n\leftrightarrow a_p}]
\end{equation}

$b^n_l(n+l)=l(-\partial_{a_l})(U^n_l)Str\left(\ldots \phi^{j_l}\ldots \right)-k(l(A_1+A_2)+n(B_1+B_2)) $
\begin{equation}
-n(D^n_l|_{i_n\leftrightarrow a_{l+1}}+\ldots +D^n_l|_{i_n\leftrightarrow a_p})
\end{equation}
Note that $(-\partial_{a_l})(U^n_l) Str\left(\ldots \phi^{j_l}\ldots \right)=(-\partial_{a_s}) (U^n_l) Str\left(\ldots \phi^{j_s}\ldots \right)$, for any $s=1,2,\ldots l-1$; we get a minus sign from $a_l \leftrightarrow a_s$, and another minus sign from $j_s \leftrightarrow j_l$.

Let's evaluate,  $ l(A_1+A_2)+n(B_1+B_2)= (lA_1+nB_1)+(lA_2+nB_2)$,
\begin{equation}
lA_1+nB_1= l U^n_l Str\left(\ldots \phi^{j_l}\ldots \underbrace{\partial_{a_l}\phi^{i_{2k}'}\phi^{i_{2k-1}'}}\ldots \right)+n U^n_l Str\left(\ldots \phi^{j_l}\ldots \underbrace{\phi^{i_n}\phi^{i_{2k-1}'}}\ldots \partial_{a_l}\phi^{i_{2k}'}\ldots \right)
\end{equation}
\begin{equation}
lA_1+nB_1= l U^n_l Str\left(\ldots \phi^{i_{2k-1}'}\ldots \underbrace{\partial_{a_l}\phi^{j_l}\phi^{i_{2k}'}}\ldots \right)+n U^n_l Str\left(\ldots \phi^{i_{2k-1}'}\ldots \underbrace{\phi^{i_n}\phi^{i_{2k}'}}\ldots \partial_{a_l}\phi^{j_l}\ldots \right)
\end{equation}
 All the $Str\left(\ldots\right)$ terms are multiplied by $U^n_l=\frac{i^kp!}{k!n!l!(p-l)!}\left(\partial_{x^{i_1}}\ldots \partial_{x^{i_n}}\right) C^0_{i_1'\ldots i_{2k}'j_1\ldots j_l[a_{l+1}\ldots a_p}$.\newline
Using the antisymmetry in the $a$'s , and the symmetries in the dummy indices of the factor outside the trace, we have,
\begin{equation}
lA_1+nB_1=(U^n_l)\sum_{q=1}^{l} Str\left(\ldots \phi^{i_{2k-1}'}\ldots \underbrace{\partial_{a_q}\phi^{j_q}\phi^{i_{2k}'}}\ldots \right)+(U^n_l)\sum_{s=1}^{n} Str\left(\ldots \phi^{i_{2k-1}'}\ldots \underbrace{\phi^{i_s}\phi^{i_{2k}'}}\ldots \right)
\end{equation}

Let's denote by $ Str\left(\ldots \underbrace{\phi^{i_{2k-1}'}}\ldots \underbrace{\phi^{i_{2k}'}}\ldots \right) $, the expression in which $\phi^{i_{2k-1}'}$ and $\phi^{i_{2k}'}$ are distinct entries in the symmetrized trace prescription, without any constraint on the "left neighbour" of  $\phi^{i_{2k}'}$. However, the "left neighbour" can only be one of the following: $\partial_{a_q}\phi^{j_q}|_{q=1,\ldots ,l}$ , or $\phi^{i_s}|_{s=1,\ldots ,n}$ , or $(\phi^{i_{2j}'}\phi^{i_{2j-1}'})|_{j=1,\ldots k-1},$   or $\phi^{i_{2k-1}'}$. 

$ lA_1+nB_1=(n+l+k)U^n_l Str\left(\ldots \underbrace{\phi^{i_{2k-1}'}}\ldots \underbrace{{\phi^{i_{2k}'}}}\ldots \right) - $
\begin{equation}
-\sum_{j}U^n_l Str\left(\ldots \phi^{i_{2k-1}'}\ldots \underbrace{\phi^{i_{2j}'}\phi^{i_{2j-1}'}\phi^{i_{2k}'}}\ldots \right)-U^n_l Str\left(\ldots \underbrace{\phi^{i_{2k-1}'}\phi^{i_{2k}'}}\ldots \right)
\end{equation}
One can notice that the last term is $b^n_l$, while the first term is 0, since $U^n_l$ is antisymmetric in $i_{2k-1}'$ and $i_{2k}'$. 
\begin{equation}
lA_1+nB_1=b^n_l -(k-1)U^n_l Str\left(\ldots \phi^{i_{2k-1}'}\ldots \underbrace{\phi^{i_{2j}'}\phi^{i_{2j-1}'}\phi^{i_{2k}'}}\ldots \right).
\end{equation}

    We can  ilustrate the type of identity that we used with a concrete  example:

$\hspace{1cm}Str\left(XY\underbrace{ZT}\right)=3Str\left(XYZT\right)-Str\left(XZ\underbrace{YT}\right)-Str\left(YZ\underbrace{XT}\right),$ \newline
where $X,Y,Z,T$ are some $N\times N$ matrices.
\newline
Similarly for,
\begin{equation}
nB_2+lA_2= n U^n_l Str\left(\ldots \phi^{i_{2k}'}\ldots \underbrace{\phi^{i_{2k-1}'}\phi^{i_n}}\ldots \partial_{a_l}\phi^{j_l}\ldots \right) +l U^n_l Str\left(\ldots \phi^{i_{2k}'}\ldots \underbrace{\phi^{i_{2k-1}'}\partial_{a_l}\phi^{j_l}}\ldots \right)
\end{equation}

$nB_2+lA_2=(n+l+k)U^n_l Str\left(\ldots \underbrace{\phi^{i_{2k}'}}\ldots \underbrace{\phi^{i_{2k-1}'}}\ldots \right)$
\begin{equation}
-(k-1) U^n_l Str\left(\ldots \underbrace{\phi^{i_{2k-1}'}\phi^{i_{2j}'}\phi^{i_{2j-1}'}}\ldots \right)- U^n_l Str\left(\ldots \underbrace{\phi^{i_{2k-1}'}\phi^{i_{2k}'}}\ldots \right)
\end{equation}
\begin{equation}
nB_2+lA_2= b^n_l-(k-1)U^n_l Str\left(\ldots \phi^{i_{2k}'}\ldots \underbrace{\phi^{i_{2j}'}\phi^{i_{2j-1}'}\phi^{i_{2k-1}'}}\ldots \right)
\end{equation}
From equations (18) and (21), we get:
\begin{equation}
l(A_1+A_2)+n(B_1+B_2)= 2b^n_l
\end{equation}
Then equation (13)  gives,
\begin{equation}
b^n_l=\frac{l}{n+l+2k}(-\partial_{a_l})(U^n_l)Str\left(\ldots \phi^{j_l}\ldots \right)-\frac{n}{n+l+2k}\left[D^n_l|_{i_n\leftrightarrow a_{l+1}}+\ldots +D^n_l|_{i_n\leftrightarrow a_p}\right]. 
\end{equation}

Let's remind ourselves what these terms really are,
\begin{equation}
\begin{split}
(-\partial_{a_l})(U^n_l)Str\left(\ldots \phi^{j_l}\ldots \right)&=\frac{-i^kp!}{k!n!l!(p-l)!}\left(\partial_{x^{i_1}}\ldots \partial_{x^{i_n}}\partial_{a_l}\right) C^0_{i_1'\ldots i_{2k}'j_1\ldots j_la_{l+1}\ldots a_p} \\
  &  Str\left(\partial_{a_1}\phi^{j_1}\ldots \partial_{a_{l-1}}\phi^{j_{l-1}}\underline{\phi}^{j_l}\phi^{i_1}\ldots \phi^{i_n}\underbrace{\phi^{i_{2k}'}\phi^{i_{2k-1}'}}\ldots \right)
\end{split} 
\end{equation}

\begin{equation}
\begin{split}
D^n_l|_{i_n\leftrightarrow a_{l+1}} &=\frac{i^kp!}{k!n!l!(p-l)!}\left(\partial_{x^{i_1}}\ldots \partial_{x^{i_{n-1}}}\partial_{a_{l+1}}\right) C^0_{i_1'\ldots i_{2k}'j_1\ldots j_li_na_{l+2}\ldots a_p} \\
 &Str\left(\partial_{a_1}\phi^{j_1}\ldots \partial_{a_{l-1}}\phi^{j_{l-1}}\underline{\phi}^{j_l}\phi^{i_1}\ldots \phi^{i_{n-1}}\partial_{a_l}\phi^{i_n}\underbrace{\phi^{i_{2k}'}\phi^{i_{2k-1}'}}\ldots \right)
\end{split}
\end{equation}

If we use the same expansion for $b^{n-1}_{l+1}$, we get,
\begin{center}
$b^{n-1}_{l+1}=\frac{l+1}{n+l+2k}(-\partial_{a_{l+1}})(U^{n-1}_{l+1})Str\left(\ldots \phi^{i_{n+l+1}}\ldots \right)$
\begin{equation}
-\frac{n-1}{n+l+2k}\left[D^{n-1}_{l+1}|_{i_n\leftrightarrow a_{l+2}}+\ldots +D^{n-1}_{l+1}|_{i_n\leftrightarrow a_p}\right].
\end{equation}
\end{center}
We can see that the second  term in the expression for $b^n_l$ is the same as the first in the expression for $b^{n-1}_{l+1}$. All the $D^n_l|_{(\ldots )}$ terms from (23) are equal to each other, due to the antisymmetry in the $a$'s, and as it turns out, they come with the right sign to cancel the first  term from (26). Let's check the numerical coefficients,

second term of $b^n_l$: $\frac{p!}{k!n!l!(p-l)!}\frac{n}{n+l+2k}(p-l) $\newline
Note that this is 0, for $l=l_{max}=p$, or for $n=n_{min}=0$. 

first term of $b^{n-1}_{l+1}$:  $\frac{p!}{k!(n-1)!(l+1)!(p-l-1)!}\frac{l+1}{n+l+2k}$ \\
The numerical factors are the same so all the terms cancel against each other, except for the $b^n_0$, and the first term of $b^n_1$ given by equation (23). 
The  $l=0$ case has to be analysed separately, since we cannot integrate by parts in this case.

\begin{equation}
b^n_0= \frac{1}{k!n!}(\partial_{x^{i_1}}\ldots \partial_{x^{i_n}})C^0_{i_1'\ldots i_{2k}'a_1\ldots a_p}Str\left(\phi^{i_1}\ldots \phi^{i_n}\ldots \underbrace{\phi^{i_{2k}'}\phi^{i_{2k-1}'}}\ldots \right).
\end{equation}

As in  equation (17), we can write:

$0=(n+k) U^n_0 Str\left(\ldots \underbrace{\phi^{i_{2k}'}}\ldots \underbrace{\phi^{i_{2k-1}'}}\ldots \right)=nU^n_0Str\left(\ldots \underbrace{\phi^{i_n}\phi^{i_{2k-1}'}}\ldots \right)+$
\begin{equation}
(k-1)U^n_0Str\left(\ldots \underbrace{\phi^{i_{2j}'}\phi^{i_{2j-1}'}\phi^{i_{2k-1}'}}\ldots \right)+U^n_0Str\left(\ldots \underbrace{\phi^{i_{2k}'}\phi^{i_{2k-1}'}}\ldots \right)
\end{equation}
Similarly,

$0=n U^n_0Str\left(\ldots \underbrace{\phi^{i_{2k}'}\phi^{i_n}}\ldots \right)+(k-1)U^n_0Str\left(\ldots \underbrace{\phi^{i_{2k}'}\phi^{i_{2j}'}\phi^{i_{2j-1}'}}\ldots \right)$
\begin{equation}
+U^n_0Str\left(\ldots \underbrace{\phi^{i_{2k}'}\phi^{i_{2k-1}'}}\ldots \right)
\end{equation}
From (28) and (29) we have:

$2U^n_0Str\left(\ldots \underbrace{\phi^{i_{2k}'}\phi^{i_{2k-1}'}}\ldots \right)=-nU^n_0Str\left(\ldots \underbrace{\phi^{i_n}\phi^{i_{2k-1}'}}\ldots \right)-n U^n_0Str\left(\ldots \underbrace{\phi^{i_{2k}'}\phi^{i_n}}\ldots \right)=$
\begin{equation}
=-n( U^n_0|_{i_n\leftrightarrow i_{2k}'}+U^n_0|_{i_n\leftrightarrow i_{2k-1}'})Str\left(\ldots \underbrace{\phi^{i_{2k}'}\phi^{i_{2k-1}'}}\ldots \right)
\end{equation}
Repeating this $k$ times, we end up with:
\begin{equation}
U^n_0Str\left(\ldots \underbrace{\phi^{i_{2k}'}\phi^{i_{2k-1}'}}\ldots \right)=\frac{n}{n+2k}(U^n_0-U^n_0|_{i_n\leftrightarrow i_1'}-\ldots -U^n_0|_{i_n\leftrightarrow i_{2k}'})Str\left(\ldots \underbrace{\phi^{i_{2k}'}\phi^{i_{2k-1}'}}\ldots \right).
\end{equation}  

From (23) we find that the first term of $b^{n-1}_1$ is (rename $j_1\rightarrow i_n$):
\begin{equation}
-\frac{i^k p}{k!(n-1)!(n+2k)} \partial_{x^{i_1}}\ldots \partial_{x^{i_{n-1}}}\partial_{a_1} C^0_{i_1'\ldots i_{2k}'{i_n}a_2\ldots a_p}Str\left(\phi^{i_1}\ldots \phi^{i_n}\ldots \underbrace{\phi^{i_{2k}'}\phi^{i_{2k-1}'}}\ldots \right)
\end{equation}
 Now we can sum the $b^n_0$ term and the first term of $b^{n-1}_1$ to get a gauge invariant term equal to:
\begin{center}
$ \frac{i^k}{k!(n-1)!(n+2k)}\partial_{x^{i_1}}\ldots \partial_{x^{i_{n-1}}} F^{0,(2k+p+1)}_{i_1i_1'\ldots i_{2k}'a_1\ldots a_p}Str\left(\phi^{i_1}\ldots \phi^{i_n}\underbrace{\phi^{i_{2k}'}\phi^{i_{2k-1}'}} \ldots \right)$, 
\end{center}
where  $F^{0,(2k+p+1)}= dC^{0,(2k+p)}$.  The monopole coupling doesn't show up in the previous expression, since in deriving equation (31) we assumed $k>0$.
Keeping track of the gauge invariant terms dropped in equation (11), we can express the total coupling, for $F=0$, as:\newline

$\mu_{p-1}\sum_{l,n>0}\frac{\lambda^{(k+n+l)}i^kp!}{k!(n-1)!l!(p-l)!(2k+n+l)}$
\begin{equation}
\partial_{x^{i_1}}\ldots \partial_{x^{i_{n-1}}} F^{0,(2k+p+1)}_{i_1i_1'\ldots i_{2k}'j_1\ldots j_la_{l+1}\ldots a_p}Str\left(\partial_{a_1}\phi^{j_1}\ldots \partial_{a_l}\phi^{j_l}\phi^{i_1}\ldots \phi^{i_n}\phi^{i_{2k}'}\phi^{i_{2k-1}'}\ldots \right). 
\end{equation}
For $k=0$ we need to add the usual monopole coupling given by $\mu_{p-1}C^0_{a_1\ldots a_p}$ from (27).

\section{ Gauge invariance for $F\neq 0$}
For non-zero $F$ along the brane, the pull-back is defined using covariant derivatives.  There are a few useful relations involving covariant derivatives that  allow us to use the previous proof in the case when $F$ is nonzero.
If $Y,Y_1,Y_2$ are $N\times N$ matrices transforming in the adjoint representation of the gauge group  ($D_aY=\partial_aY+i[A_a,Y]$), and $f$ is a scalar function, then:

(a) $Tr[D_a(Y_1Y_2)]= Tr[D_a(Y_1) Y_2]+Tr[Y_1(D_aY_2)]$

(b) $D_a(fY)=(\partial_af)Y+fD_aY$

(c) $[D_1,D_2]Y = i [ F_{12},Y]$, where  $F_{12}=\partial_1A_2-\partial_2A_1+i[A_1,A_2]$

(d) $D_{[a}F_{bc]}=0$, by the Bianchi identity

  In this case, the equivalent of equation (6), which gives
the coupling   between $N$ $D_{p-1}$ branes and $C_{p+2k}$ is:

$ 
\sum_{l,n,r} \frac{\mu_{p-1}\lambda^{k+n+l+2r}i^{k+r}p!}{2^rr!(k+r)!(n)!l!(p-l-2r)!}\partial_{x^{i_1}}\ldots \partial_{x^{i_n}} C^0_{i_1'\ldots i_{2(k+r)}'j_1\ldots j_l[a_{l+1}\ldots a_{p-2r}} $
\begin{equation}
 Str \left(D_{a_1}\phi^{j_1}\ldots D_{a_l}\phi^{j_l}\phi^{i_1}\ldots \phi^{i_n}\underbrace{\phi^{i_{2(k+r)}'}\phi^{i_{2(k+r)-1}'}}\dots \underbrace{\phi^{i_2'}\phi^{i_1'}}\ldots F_{a_pa_{p-1}]} \right),
\end{equation}
where  $ 0\le l\le p-2r$, and $r$ is the number of $F$'s appearing inside the Str part. 
As in the proof for $F=0$, we will write the sum in (34) as  $\mu_{p-1}\sum_{l,n,r}\lambda^{k+n+l+2r}b^n_{l,r}$, and denote by $U^n_{l,r}$ the term outside the trace corresponding to $b^n_{l,r}.$
When integrating  $b^n_{l,r}$ by parts,  now we will have extra  terms containing $Str\left(\ldots D_{[a_1}D_{a_2]}\phi\ldots \right)$. Since $D_{[a_1}D_{a_2]}\phi=\frac{1}{2}[D_{a_1}, D_{a_2}]\phi=\frac{i}{2}[F_{a_1a_2},\phi]$  these extra terms will cancel against other terms in the expansion for $b^n_{l-2,r+1}$. Given these facts, the right-hand side of equation (7) has an aditional term equal to:
\begin{center}
$(l-1)(-U^n_{l,r})Str\left(\phi^{j_1}\ldots D_{a_1}D_{a_l}\phi^{j_l}\ldots \right)$,
\end{center}
while (18) changes to: 
\begin{center}
$lA_1+nB_1=b^n_{l,r} -(k+r-1)U^n_{l,r} Str\left(\ldots \phi^{i_{2(k+r)-1}'}\ldots \underbrace{\phi^{i_{2j}'}\phi^{i_{2j-1}'}\phi^{i_{2(k+r)}'}}\ldots \right)-rU^n_{l,r}Str\left(\ldots \underbrace{F\phi^{i_{2(k+r)}'}}\ldots \right)$.
\end{center}
Now, we can see that the generalizations of equations (22) and (23) are:
\begin{equation}
l(A_1+A_2)+n(B_1+B_2)= 2 b^n_{l,r}-r U^n_{l,r}Str\left(\ldots \underbrace{[F,\phi^{i_{2(k+r)}'}]}\ldots \right)
\end{equation}

\begin{center}
$b^n_{l,r}=\frac{l}{n+l+2(k+r)}(-\partial_{a_l})(U^n_{l,r})Str\left(\ldots \phi^{j_l}\ldots \right)-\frac{l(l-1)}{n+l+2(k+r)}(U^n_{l,r})Str\left(\phi^{j_1}\ldots D_{a_1}D_{a_l}\phi^{j_l}\ldots \right)$ 
\end{center}
\begin{center}
$-\frac{n }{n+l+2(k+r)}\left[D^n_{l,r}|i_n\leftrightarrow a_{l+1}+\ldots +D^n_{l,r}|{i_n\leftrightarrow a_{p-2r}}\right]$
\begin{equation}
+\frac{r(k+r)}{n+l+2(k+r)}(U^n_{l,r})Str\left(\ldots \underbrace{ [F,\phi^{i_{2(k+r)}'}]}\ldots \right)
\end{equation}
\end{center}
 The second term in the expansion for $b^n_{l,r}$ has the same structure as the fourth term in the expansion for $b^n_{l-2,r+1}$.
While it is easy to see that these extra terms have the required form to produce the (partial) cancelation between $b^n_{l,r}$, and  $b^n_{l-2,r+1}$ we have to make sure the numerical pre-factors are equal:
\begin{center}
from $b^n_{l,r}$: $\frac{i^{k+r}p!}{2^r(k+r)!r!(n)!l!(p-l-2r)!}l(l-1)\frac{i}{2}$, \newline

from $b^n_{l-2,r+1}$: $\frac{i^{k+r+1}p!}{2^{r+1}(k+r+1)!(r+1)!n!(l-2)!(p-l-2r)!}(r+1)(k+r+1)$
\end{center}
 Since the numerical factors are the same, when we are summing the $b^n_{l,r}$'s over $r$,  all the extra terms that we get in the case of a non-zero $F$  will cancel against each other, except for the second term of $b^n_{l=2,r}$.   At the limits, when $r=r_{min}$ the fourth term in the expression for $b^n_{l,r}$ is 0, since $r_{min}(r_{min}+k)=0$. If $k>0$, $r_{min}=0$, otherwise $r_{min}=-k$. When $r=r_{max}$, $l<2$ so the second term in the expansion for $b^n_{l,r}$ is 0. 
After summing over $l$ and $r$ we are left with:
\begin{center}
$\sum_r(b^n_{l=0,r})+\sum_r\frac{1}{n+2(k+r)}(-\partial_{a_1})(U^{n-1}_{l=1,r})Str\left(\ldots \phi^{j_1}\ldots \right)$
\begin{equation}
+\sum_{r>r_{min}}\frac{-i}{n+2(k+r)}U^n_{2,r-1}Str\left(\ldots \phi^{j_1}\ldots \underbrace{[F_{a_1a_l},\phi^{j_l}]}\ldots \right) 
\end{equation}
\end{center}
For $b_{l=0,r}$ we are using a transformation as in (31):
\begin{center}
$U^n_{0,r}Str\left(\ldots \underbrace{\phi^{i_{2(k+r)}'}\phi^{i_{2(k+r)-1}'}}\ldots \right)=\frac{n}{n+2(k+r)}(U^n_{0,r}-U^n_{0,r}|_{i_n\leftrightarrow i_1'}-\ldots -U^n_{0,r}|_{i_n\leftrightarrow i_{2(k+r)}'})Str\left(\ldots \underbrace{\phi^{i_{2(k+r)}'}\phi^{i_{2(k+r)-1}'}}\ldots \right)$
\begin{equation}
-\frac{(k+r)(r)}{n+2(k+r)}U^n_{0,r}Str\left(\ldots \underbrace{[F,\phi^{i_{2(k+r)-1}'}]}\ldots \right)
\end{equation} 
\end{center}

Using  (38) to replace $b^n_{l=0,r}$ in (37), we get:
\begin{center}
$\sum_r\frac{n}{n+2(k+r)}(U^n_{0,r}-U^n_{0,r}|_{i_n\leftrightarrow i_1'}-\ldots -U^n_{0,r}|_{i_n\leftrightarrow i_{2(k+r)}'})Str\left(\ldots \underbrace{\phi^{i_{2(k+r)}'}\phi^{i_{2(k+r)-1}'}}\ldots \right)$ \newline
$+\sum_r\frac{1}{n+2(k+r)}(-\partial_{a_1})(U^{n-1}_{l=1,r})Str\left(\ldots \phi^{j_1}\ldots \right)$.
\end{center}
The above expression is gauge invariant since we can write it as a field strength term noticing that after renaming $j_1\rightarrow i_n$, the $Str$ parts are identical and:
\begin{center}
$(\partial_{a_1})(U^{n-1}_{l=1,r})= n(p-2r)U^n_{0,r}|_{i_n\leftrightarrow a_1}$.
\end{center}
Taking into account the corresponding gauge invariant terms dropped in equation (11) the total coupling between $N$ $D_{p-1}$ branes and a $C_{p+2k}$ potential can be expressed in a gauge invariant way as:
\begin{center}
$
\mu_{p-1}\sum_{r,l}\frac{\lambda^{k+1+2r+l}i^{k+r}p!}{2^r r!(k+r)!l!(p-2r-l)!}
Str(\overline{F}_{r,l}^{(2k+p+1)}(\phi)_{i_1i_1'i_2'\ldots i_{2(k+r)-1}'i_{2(k+r)}'j_1\ldots j_la_{l+1}\ldots a_{p-2r}}\phi^{i_1}D_{a_1}\phi^{j_1}\ldots D_{a_l}\phi^{j_l}$
\begin{equation} 
\underbrace{\phi^{i_{2(k+r)}'}\phi^{i_{2(k+r)-1}'}}\ldots \underbrace{\phi^{i_2'}\phi^{i_1'}}F_{a_{p-2r+1}a_{p-2r+2}}\ldots F_{a_{p-1}a_p}).
\end{equation}
\end{center}
where we defined,\newline
\hspace*{1.5cm}$\overline{F}_{r,l}^{(2k+p+1)}(\phi)=\sum_{n\geq0} \frac {\lambda^n}{(n)!(n+l+2k+2r+1)} \phi^{i_1}\ldots \phi^{i_{n}} \partial_{x^{i_1}}\ldots \partial_{x^{i_{n}}} F^{0,(2k+p+1)}(\sigma,x^i)|_{x^i=0}.$ \newline
 Since equation (38) was derived assuming $k>0$, for $k\leq 0$  there is an additional monopole coupling term given by:
\begin{equation}
\mu_{p-1}\frac{\lambda^{|k|}p!}{2^{|k|}(|k|)!(p-2|k|)!}C^0_{[a_1\ldots a_{p-2|k|}}F\ldots F_{a_{p-1}a_p]}.
\end{equation}

\section{Discussion}

    We obtained a manifestly gauge invariant expression for the Chern-Simons coupling between $N$ $D_{p-1}$ branes and a RR potential, $C_{p+2k}$. In the presence of a 2-form $B$ field, the gauge transformations of the RR fields, become:
\begin{equation}
\sum_n C^{(n)}e^B \rightarrow \sum_n C^{(n)}e^B + d\sum_p \Lambda^{(p)}
\end{equation}
The presence of the $B$ field does not affect the generality of the previous proof since, from the point of view of the gauge transformations,  we can absorb $B$ into the definition of the RR fields. However, the proof in this note applies only for finite $N$. For $N \rightarrow \infty$ we can no longer use the property of cyclicity of the trace, and we expect monopole couplings even to higher rank RR fields. As in  matrix model,  one can construct a higher dimensional brane  out of an infinite number of lower dimensional ones, hence in (1) we should have source terms for higher dimensional D-brane charges.  

\section{Acknowledgments}

 I would like to thank   John Schwarz for suggestions  throughout this project. I also benefited from discussions with  Peter Lee, Arkadas Ozakin, Jongwong Park, and Costin Popescu.
 
 After this work was completed, the paper ~\cite{emil} appeared. It  might shed new insight on the issues we studied in this note.

\end{document}